# Exploring Broad-Spectrum Antimicrobial Nanotopographies: Implications for Bactericidal, Antifungal, and Virucidal Surface Design

Vladimir A. Baulin[a], Denver P. Linklater[b], Saulius Juodkazis[c], Elena P. Ivanova[d*]

[a]Universitat Rovira i Virgili, Departament d'Enginyeria Quimica, Tarragona 43007, Spain

[b]Department of Biomedical Engineering, Graeme Clark Institute, The University of Melbourne, Parkville, Victoria 3010, Australia

[c]Swinburne University of Technology, Hawthorn, Victoria 3021, Australia

[d]School of Engineering, STEM College, RMIT University, Melbourne, Victoria 3000, Australia

Corresponding author email: elena.ivanova@rmit.edu.au

**Abstract**

Inspired by the natural defence strategies of insect wings and plant leaves, nanostructured surfaces have emerged as a promising tool in various fields, including engineering, biomedical sciences, and materials science to combat bacterial contamination and disrupt biofilm formation. However, the development of effective antimicrobial surfaces against fungal and viral pathogens presents distinct challenges, necessitating tailored approaches. Here, we aimed to review the recent advancements of the use of nanostructured surfaces to combat microbial contamination, particularly focusing on their mechanobactericidal and antifungal properties, as well as their potential in mitigating viral transmission. We comparatively analysed the diverse geometries and nano-architectures of these surfaces and discussed their applications in various biomedical contexts, such as dental and orthopedic implants, drug delivery systems, and tissue engineering. Our review highlights the importance of preventing microbial attachment and biofilm formation, especially in the context of rising antimicrobial resistance and the economic impact of biofilms. We also discussed the latest progress in material science, particularly nanostructured surface engineering, as promising strategies for reducing viral transmission through surfaces. Overall, our findings underscore the significance of innovative strategies to mitigate microbial attachment and surface-mediated transmission, while also emphasizing the need for further interdisciplinary research in this area to optimize antimicrobial efficacy and address emerging challenges.

**Vocabulary**

**Fungicidal:** Able to kill fungal pathogens

**Virucidal:** having the capacity to or tending to destroy or inactivate viruses

**Mechano-bactericidal:** The action of natural or synthetic nanomaterials that can physically induce bacterial cell death through the application of physical forces.

**Micro–nanostructuring:** A technique used to impart physical patterns onto a surface with surface features (protrusions or depressions) on the scale of micrometres or nanometres. The size and shapes of such patterns determine the overall topography of the surface at these scales. Such patterns can be made by depositing material onto a surface, removing material from a surface, or both.

**Nanotopography:** The specific surface features that form or are generated at the nanoscale

**Stretching modulus:** Also known as Young's modulus. The modulus of elasticity: a measure of the ability of a material to withstand changes in length when under lengthwise tension or compression.

**Biomimetic:** Imitating the models, systems and elements found in nature.

1. **INTRODUCTION**

    **1.1 Foundation and Multifunctionality of Nanostructured Surfaces**

    The nanoengineering of surfaces has led to significant technological advances in various fields, including engineering[1], biomedical sciences[2, 3], and materials science[4]. Nanostructured surfaces possess unique properties owing to their large surface area, tunable porosity, and the possibility of surface functionalization[5, 6]. The diversity of nanostructures that can be fabricated on a plethora of materials endows surfaces with an assortment of functional capabilities (Figure 1). In applications such as energy harvesting, catalysis, and optics, nanotopography facilitates control over phenomena occurring at the material interface. For instance, surface nanotopography can influence wettability, optical reflectivity, absorption[7], and electrical conductivity[8], crucial for specific chemical reactions or physical interactions with the environment[9]. From a mechanical perspective, surface nanoroughness influences friction, wear, and lubrication,[10] affecting the movement of components against each other[11]. On a broader scale, surface nanotopography can influence thermal properties by altering heat transfer and distribution. Indeed, in heat exchanger applications, rough surfaces can disrupt laminar flow and create turbulence, which may enhance the heat transfer efficiency[12]. In biomedical applications, surface nanostructuring has been used to enhance the efficacy of drug delivery systems, bioimaging, biosensing, and tissue engineering, and has received particular attention in the development of antimicrobial biomaterials, such as dental and orthopedic implants[5, 13, 14].

    Biomimetic nanostructured surfaces, inspired by naturally occurring patterns like those found on the intricately structured surfaces of insect wings[15, 16], eyes[7, 17, 18], and plant surfaces demonstrate effective microbial eradication strategies[19, 20]. Approaches to replicate these nanostructures from nature have the potential to revolutionize our strategies against microbial threats. These nanoscale topographical features exert

lethal effects on bacterial cells through physical mechanisms, such as mechanical rupture upon contact, thereby disrupting the initial stages of biofilm formation[21]. One prominent example of the first mechano-bactericidal nanopattern found in nature is on the wings of the cicada *Psaltoda claripennis*[22]. The physical nature of the mechano-bactericidal action has a notable advantage over chemical disinfection methods by circumventing potential resistance mechanisms that microorganisms may develop over time[13, 23-25].

Nanostructured antifungal surfaces *are distinct* from their antibacterial counterparts and designed specifically to address the challenges posed by fungi, which exhibit structural and functional differences from bacteria[26]. Fungi, ranging from unicellular yeasts to filamentous molds, apart from significantly larger sizes, possess a higher level of complexity in their cellular structure, cell envelope, and lifestyle than bacteria. Their robust cell wall, composed of chitin, glucans, and proteins, form an efficient protective barrier to external mechanical stressors[26, 27]. Additionally, fungi display a range of morphotypes, such as spores and hyphae, which enhance their virulence and ability to penetrate substrata[28, 29]. Thus, the geometry of the nanopattern must be meticulously engineered to prevent spore germination, impede hyphal proliferation, and disrupt the fungal life cycle.

Recently, nanostructured surfaces have been revealed to possess antiviral properties. Similar to the mechanical rupture of bacterial cell walls as they adhere to high aspect ratio nanostructures of a nanopatterned surface, the lipid envelope of viruses can be punctured, or ruptured according to a similar biophysical mechanism[30]. However, just as antifungal surfaces require a tailored approach to create a surface that can overcome the physical robustness of fungi and yeasts, antiviral surfaces must be designed specifically to target the nanoscale size of virus particles. Indeed, viruses are one order of magnitude smaller than bacteria or fungi.

## 1.2 Importance of Preventing Microbial Attachment and Biofilm Formation on Surfaces in the Era of Antibiotic Resistance

Microbial attachment and subsequent biofilm formation present considerable challenges in medical devices, industrial equipment, and other critical interfaces (Figure 1). With a global economic impact exceeding $5,000 billion annually, biofilms affect agriculture, healthcare, food processing, industrial manufacturing, marine industries, and sanitation[31, 32]. Successful bacterial attachment in industrial systems leads to biofilm formation, causing biofouling. This process deteriorates materials, reduces heat exchanger efficiency, clogs water systems, and promotes corrosion[33]. In water treatment facilities, biofilms can harbor pathogenic organisms, thus compromising drinking water safety[34].

The economic and health-related effects of biofilms necessitate the development of innovative strategies to mitigate microbial attachment. Current approaches include antimicrobial coatings incorporating chemical agents[35], such as antimicrobial peptides[36], antibiotics targeting protein synthesis or cell wall assembly, disinfectants, antiseptics disrupting bacterial membranes or essential enzymes, and quorum-sensing inhibitors to disrupt cell-to-cell communication within biofilms[37]. In addition to traditional chemical entities, the antimicrobial arsenal is fortified by metal ions and nanoparticles[38]. Materials, such as silver and copper, disrupt cellular machinery and induce oxidative stress, leading to cell death. Nanoparticles composed of these metals or other compounds like zinc oxide or titanium dioxide utilize their unique surface properties and quantum effects to exert potent bactericidal action due to their diminutive size and extensive surface area[39]. However, these substances are increasingly ineffective against drug-resistant bacteria and yeasts[40], necessitating the exploration of novel antimicrobial approaches[19]. Contemporary therapies, such as photodynamic and photothermal methods, utilize light. These techniques employ photosensitive agents that generate lethal reactive oxygen species or materials that convert light energy into heat upon illumination, both effectively targeting bacteria[41]. For example, antibacterial textiles are widely used in peoples' practical lives. In-situ deposition or growth of diverse nanoparticles with antibacterial effects can be realized on the surface of the fibre surfaces of textiles[42]. In addition, photothermal and electrothermal nanomaterials such as MXene nanosheets, silver nano wires and other nanoparticles are commonly used in the preparation of antibacterial textiles[43].

In recent years, overuse of antibiotics and antifungals in healthcare and agriculture has fuelled antimicrobial resistance (AMR), posing a grave threat to effective bacterial infection treatment. The decline in new antibiotic development[44], paired with the rise of "superbugs" resistant to all available antibiotics, exacerbates this global issue[31]. For example, the attachment of bacteria to implant surfaces can lead to persistent infections that are difficult to eradicate because of the enhanced antibiotic resistance of biofilm-associated cells[45]. Such infections lead to device failure, increased patient morbidity, prolonged hospital stay, and even mortality in severe cases[34].

Dominance over the surface is a crucial factor toward success in the ongoing battle between host cells and invading pathogens after the implantation of a prosthetic or medical device. Surfaces that promote eukaryotic cell growth while deterring bacterial colonisation represent a dual-functional approach crucial for medical implants[46]. These materials support host cells in winning the "race for the surface," thereby favouring tissue integration and preventing infection. Surface patterns that mimic the mechano-bactericidal properties of cicada wings enhance biomaterial integration into the host without adverse immune reactions[19]. Mechano-biocidal nanostructured surfaces selectively target bacterial cells due to their size disparity with mammalian cells, ensuring minimal damage by the latter. This selectivity is crucial for maintaining biocompatibility in medical contexts. Therefore, a material's ability to resist bacterial colonization is critical in determining its suitability for biomedical applications, such as prosthetic devices, catheters, and surgical instruments.

Fungal contamination can also have severe implications for the medical, agricultural, and food industry[28, 47]. Pathogenic yeasts, such as *Candida* spp., account for an increasing percentage of implant-associated infections. They have been isolated from biofilms that colonize central venous catheters, urinary catheters, hip and knee implants, and dentures. These infections are of particular concern because of their severity and associated high fatality rates. We recently reviewed the significance of *Candida*-implant-associated infections[48]. Filamentous fungal species, such as *Aspergillus* spp., are opportunistic pathogens that pose a considerable health risk to immunocompromised individuals[29]. *Aspergillus* spp. has been implicated in oral cavity aspergillosis post-tooth extraction or endodontic treatment. Azole resistance in *Aspergillus* spp. results in invasive infections with high mortality rates[49]. Furthermore, *Aspergillus* spp. colonize various bone substitutes[50].

Recent advancements in materials science, particularly in nanostructured surface engineering, have offered promising avenues for reducing viral transmission through contaminated surfaces[35, 51]. Nanostructured surfaces designed to deactivate or repel viral pathogens upon contact offer a proactive approach to controlling surface-borne infections. These surfaces function through various mechanisms, including physically disrupting viral envelopes, photocatalytically degrading viral particles, and utilizing superhydrophobic coatings to minimize surface contact and contamination[52]. Integrating nanostructured surfaces with traditional disinfection methods can substantially reduce the burden of infectious diseases by interrupting a key transmission pathway. However, the success of these interventions depends on a comprehensive understanding and optimization of the antiviral mechanisms at the nanoscale, necessitating interdisciplinary collaboration among virologists, materials scientists, and public health professionals.

In this review, we focus on nanosurfaces designed to be effective against bacteria, fungi, and viruses. Promising surface designs integrate nano and microscale topographical features inspired by naturally occurring antipathogenic surfaces, such as those found on lotus leaves and cicada wings, known for their self-cleaning and microbial-inhibiting properties. Incorporating materials with defined geometries (*e.g.*, spikes, ridges, or pillars) at the nanoscale mechanically disrupts microbial membranes upon contact, effectively targeting a broad spectrum of pathogens, including bacteria, fungi, and enveloped viruses. We provide an overview of biomimetic nanostructured surfaces (Figure 2) as compelling templates for designing novel antimicrobial materials. We explore various topographies, architectures, and geometries of nanofeatures that dictate the antibacterial, antifungal, and antiviral mechanisms associated with specific nanofeatures.

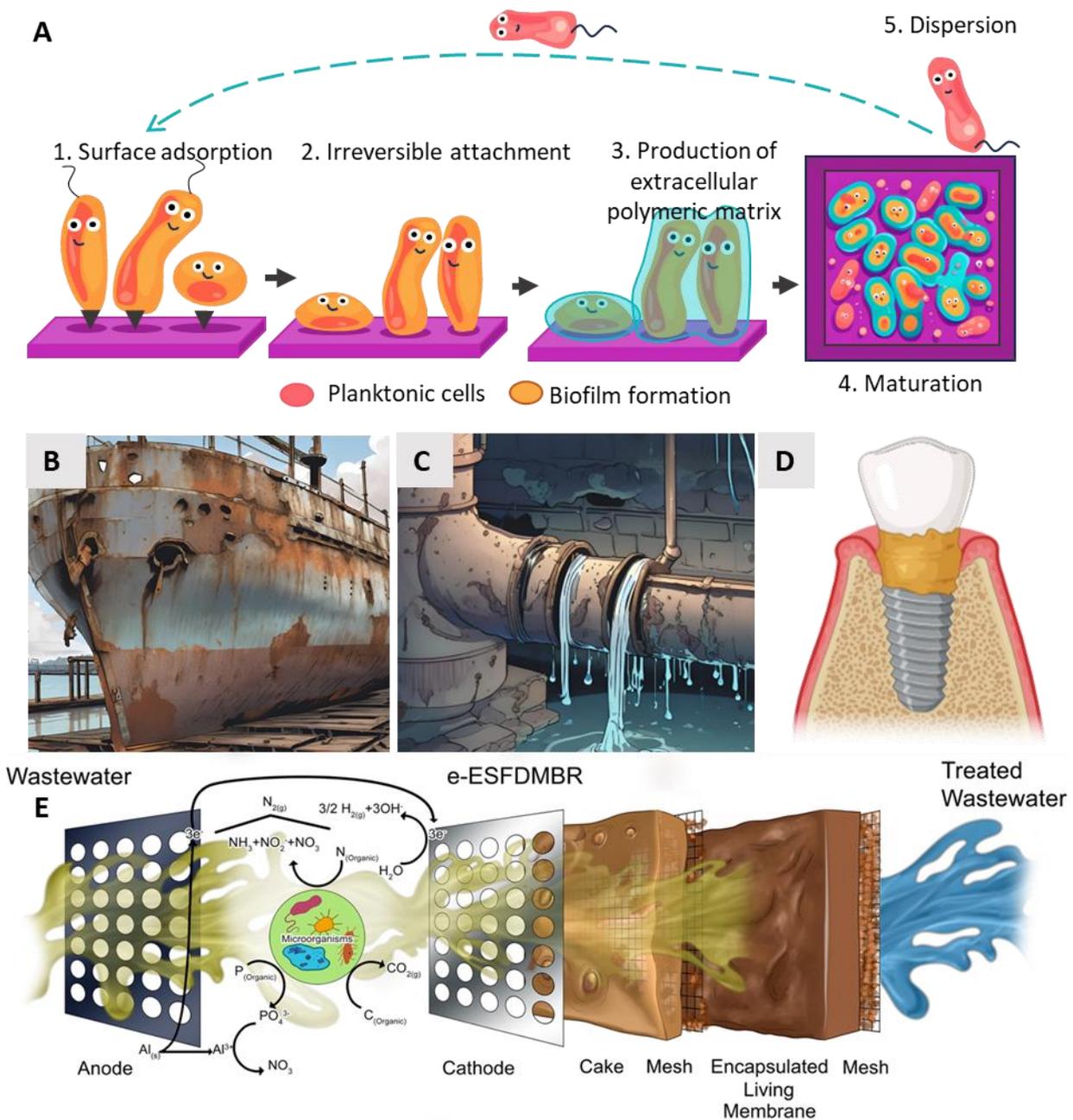

**Figure 1. Economic and health-related impacts of biofilms.** (a) Process of bacterial biofilm formation. Industries affected by microbial biofilm contamination include: (b) shipping (marine corrosion), (c) water transport (clogging, fouling), (designs b and c created using Canva.com. Copyright 2025, The Authors) (d) implantable devices (infection) (image created using BioRender.com) and (e) wastewater treatment (membrane fouling). Image reprinted with permission under a Creative Commons Attribution 4.0 International License from ref [53]. Copyright © 2022, The Author(s).

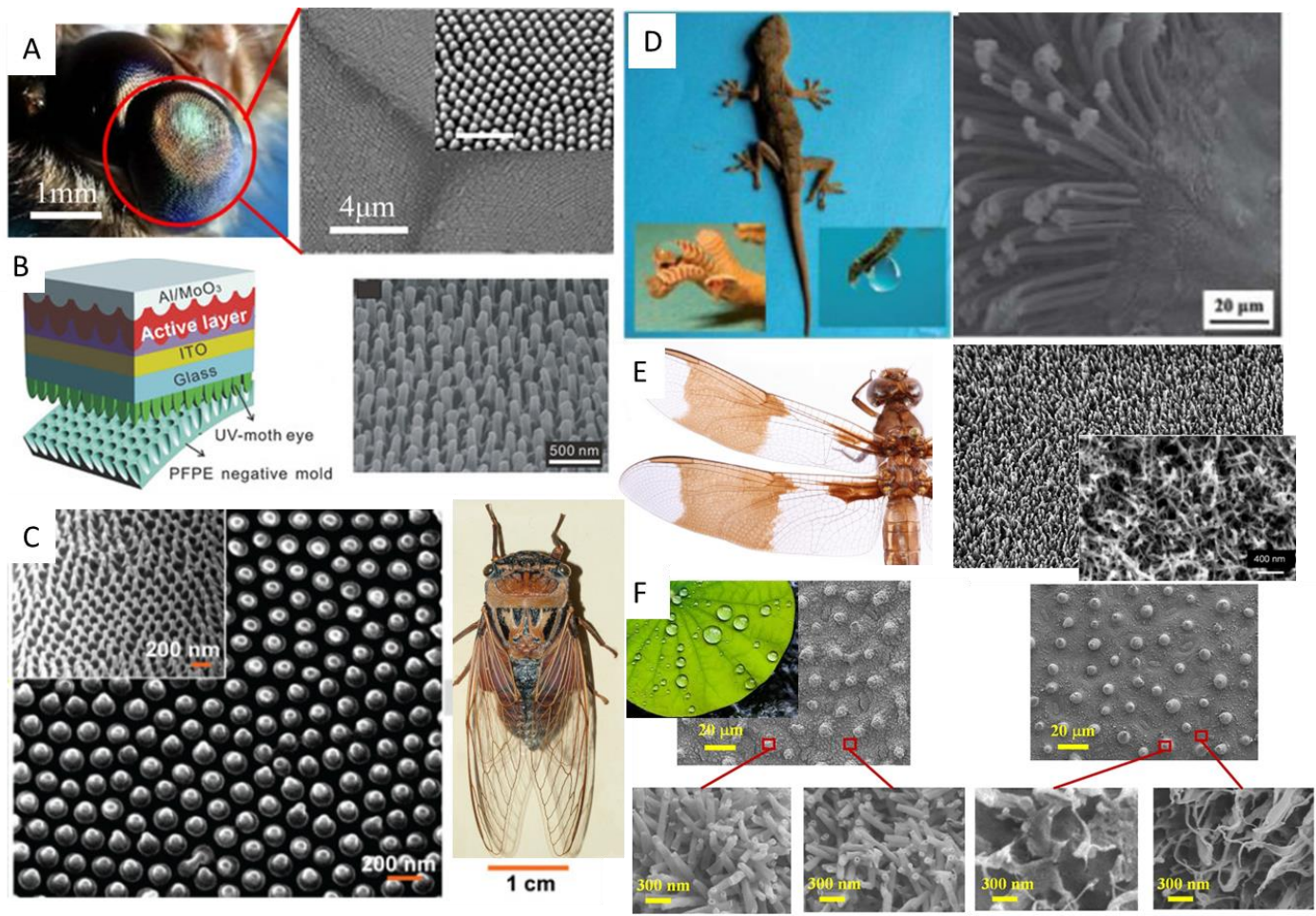

**Figure 2. Natural and biomimetic nanostructured surfaces. (a)** Surface topography of moth-eye. Image reprinted with permission under a Creative Commons Attribution 4.0 International License from ref [18]. Copyright © 2018, The Author(s). **(b)** Fabrication of biomimetic nanopattern. Image reprinted with permission under a Creative Commons Creative Commons Attribution 4.0 International License from ref [54]. Copyright © 2016 The Authors. **(c)** Nanotopography of the cicada wing. Reprinted with permission from Creative Commons Attribution 4.0 International License from ref [55]. Copyright © 2020 The Author(s). Published by Elsevier Ltd. **(d)** Superhydrophobic surface of gecko feet. Reprinted with permission under a Creative Commons Attribution 4.0 International License from ref [56]. Copyright © 2017 The Author(s). **(e)** Nanotopography of dragonfly wing and (inset) biomimetic titania surface fabricated by hydrothermal etching. Reprinted with permission under a Creative Commons Attribution 4.0 International License from ref [57]. Copyright © 2015, The Author(s). **(f)** Superhydrophobic microstructured surface of the lotus leaf and insets showing the nanostructured wax crystals and biomimetic lotus-leaf-inspired polymeric substrate with hierarchical surface topography. Adapted with permission from ref [58]. Copyright © 2019 Elsevier.

## 3. MECHANO-BACTERICIDAL ACTION OF NANOSTRUCTURED SURFACES

The interaction between microbes and surfaces is a complex process that has evolved over 3.7 billion years since bacterial existence and involves a series of events, from initial adsorption to eventual colonization.

Understanding this dynamic interplay of initial interactions of bacteria and surfaces is crucial for developing materials that resist microbial colonization and leveraging beneficial microbial interactions for environmental and biotechnological purposes[59].

The initial microbial interactions with surfaces may be influenced by microbial motility (relevant only to bacteria), surface properties, and environmental conditions[60]. Surface roughness, wettability, charge, and the presence of surface-bound molecules can either attract or repel microbial adhesion[61]. Accordingly, surfaces with specific nanostructures or chemistries have been engineered to combat microbial colonization, leveraging insights from nature and material science principles.

The random collision of microbes with a surface can be accompanied by weak, reversible adhesion driven by Van der Waals forces, hydrophobic interactions, and other noncovalent (electrostatic) interactions. Graduating from this transient adhesion phase, bacteria and fungi fortify their attachment by secreting extracellular polymeric substances. As non-living organisms, viruses do not actively colonize surfaces; however, their adsorption onto a material is crucial as a vector for enhanced disease transmission. The irreversible attachment of bacteria and fungi to a surface marks a critical shift from a planktonic or free-living state to a sessile community structure, leading to biofilm formation. Within this architecture, microbial cells are enmeshed within a self-produced extracellular polymeric substances matrix, which serves as a protective barrier against environmental threats and antimicrobial agents (Figure 1). Once anchored, microbes begin a sophisticated dialogue with their substratum and neighboring cells. They sense and respond to surface topography, stiffness, and other material cues using mechano-reception mechanisms that influence their growth, morphology, and gene expression. For example, bacterial biofilms may display heterogeneity with active growth and dormancy regions, whereas fungal hyphae can penetrate and invade surfaces, exacerbating their disruptive potential[14]. This communication within the biofilm can further lead to the dispersion of cells to colonize new surfaces (Figure 1). The broader context, including nutrient availability, pH, temperature, and the presence of other organisms, further modulates microbial surface interactions. These environmental factors can either accelerate the colonization process or inhibit microbial growth, thereby affecting both the biofilm's robustness and the onset of potential infections or material degradation. Mixed microbial biofilms resist disinfectants such as quaternary ammonium compounds and other biocides[25, 62].

We recently summarized research on the effects of surface wettability, roughness, and architecture on *Candida* spp. attachment to implantable materials and addressed nanofabrication of material surfaces as a potential method for preventing *C. albicans* spp. attachment and biofilm formation on medical implant materials[48]; and therefore these aspects are not included in this review. In the following sections, we discuss the characteristics and features of the physical rupture of bacteria by different nanopatterns.

**3.1 Antibacterial Surfaces: Lessons from Nature**

Micro and nanostructured surfaces in nature including shark skin, gecko feet, insect wings, and plant leaves, exhibit a unique surface topography that aids in antibacterial and antibiofouling capabilities[63, 64].

Natural surfaces display anti-biofouling, self-cleaning, and superhydrophobic properties due to the presence of well-ordered micro/nano-structures on their surfaces. The leaves of the taro plant, lotus plant and the petals of roses are excellent examples of superhydrophobic and antibiofouling surface topographies[3, 65, 66]; however, their superhydrophobicity is endowed by a combination of both hierarchical surface architecture on the micron-scale and waxy polygonal epidermal cells. By contrast, dragonflies and cicadas exhibit remarkable surface nanotopography on their wings[15, 22, 67]. Their wing membranes feature a dense array of nanopillars measuring between 200 and 400 nm. The nanostructures on the wings play a crucial role in determining their wettability. The wettability is affected by both the topographical features and the wax layer. Wings with high-aspect-ratio nano-protrusions arranged in an orderly fashion show greater hydrophobicity[68]. However, when submerged, this superhydrophobicity does not seem to impact the wings' ability to resist bacteria, as bacteria tend to adhere strongly to the nano-structured pillars. When bacteria encounter these nanopillars, their cell membranes stretch over them, like a balloon's skin pulled over a bed of nails. Due to the bacterial cell envelope's small size and limited mechanical strength, it breaks, causing leakage of cellular contents, including vital cytoplasmic components and nucleic acids. This loss, coupled with the ensuing imbalance between the internal and external environment of the cell, initiates a cascade of fatal events, ultimately culminating in cell death.

Remarkably, the mechanical action of nanopatterns does not depend on the material chemical nature as it was shown that the deposition of 6-10 nm thin gold films on the surface of cicadas, dragonflies, or black silicon while changing the chemistry of the surface, did not significantly changes the surface nanopattern and did not compromise their bactericidal effect[19]. This also means that surface wettability does not play a significant role in bactericidal effect, as reported elsewhere[69]. However, the geometry and topology of the nanopatterns significantly affect the selectivity and efficacy of the bactericidal effect. For example, the regular array, which was found on cicada wing surface effectively kills gram-negative bacteria only, whereas the dragonflies and damselflies nano-patterns successfully ruptured gram-negative, gram-positive bacteria and *Bacillus subtillis* spores[16, 69, 70]. This physical rupture mechanism avoids the pitfalls of chemical antimicrobial resistance, as bacterial cells are less likely to develop resistance mechanisms to mechanical disruption than against drugs or biocides. This is because cell rupture and death occur rapidly, typically within minutes, depending on the nanopattern involved[71]. Furthermore, physical cell injury has been shown to ultimately result in cell death, regardless of whether the initial mechanical injury is non-lethal[72, 73].

**3.2 Overview and Classification of Mechano-bactericidal Mechanisms**

Mechano-bactericidal mechanisms can be classified according to the nature of the physical interaction between the bacteria and the bactericidal surface, as well as the resulting morphological changes to the bacterial cell. Historically, microbial death on nanostructured surfaces has been categorized into three regimes: contact, deformation, and membrane stretching. Ivanova et al. were the first to report bacterial cell lysis on a nanopatterned substratum. They proposed that the adsorption of cells to the nanotopography of cicada wing surfaces leads to self-induced deformation and subsequent cell lysis[22]. A biophysical model was developed by Pogodin et al.[21] to evaluate the cell wall stretching mechanism on shallow cicada wing cones that cannot pierce the cell walls. Mechanical deformation and eventual cell membrane rupture by surface features were suggested to be the principal causes of bactericidal efficacy[21]. In 2020, a critical review of the mechanisms of mechano-bactericidal action of nanostructured surfaces has summarised various nanopatterns and their effects on bacterial cells[14]. Mechano-bactericidal mechanisms leverage the mechanical vulnerabilities of bacterial cell walls and membranes by different means (Figure 3). Currently, reported mechano-bactericidal mechanisms include microbial membrane stretching and rupture upon adhesion to high aspect ratio structures, membrane piercing by sharp structures, and membrane tearing due to external forces such as shear flow acting on the cell[14]. Herein, we elaborate on and extend the previously identified mechano-bactericidal actions based on a recent literature survey.

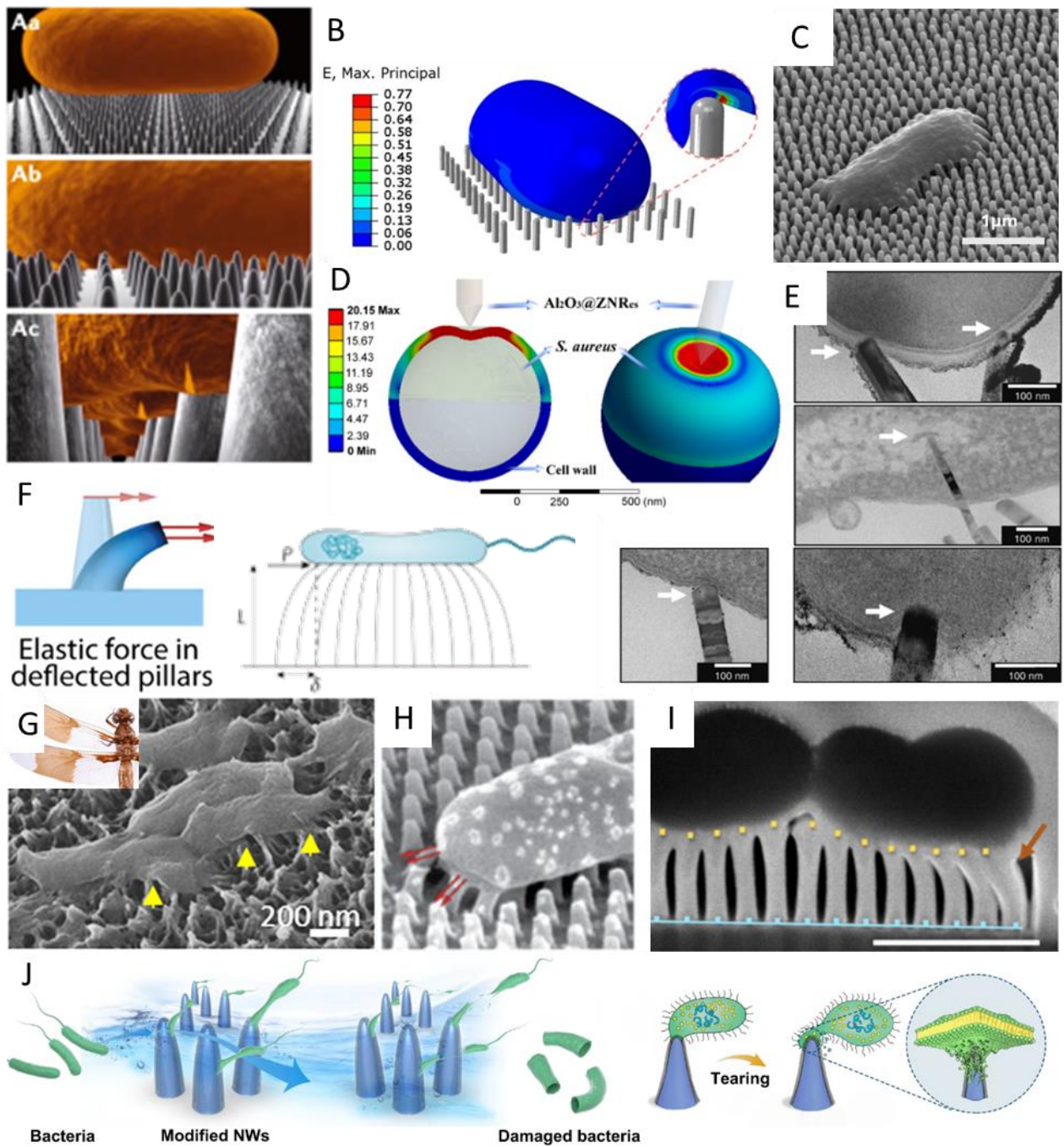

**Figure 3. The mechano-bactericidal action associated with characteristics nanopatterns**. (a) Illustration of the biophysical model produced by Pogodin *et al*., whereby bacterial cell rupture occurs by continued suspension of the bacterial cell membrane on the nanopatterned substratum, leading to stretching beyond the elastic limit of the membrane. Reprinted with permission from ref [21]. Copyright © 2013 Elsevier Inc. (b) A finite element method model showing the location of critical stress on the bacterial membrane achieved by adhesion to the cicada-like nanopatterned surface. The model incorrectly assumes the application of force (pressure) from above leading to the bacteria being unrealistically deformed. Reprinted with permission from ref [74]. Copyright © 2021 Elsevier. (c) SEM image of rod-shaped bacteria ruptured by cicada wing nanopattern. (d) The finite element method (FEM) model shows the *S. aureus* cell wall puncturing by a sharp nanopillar. The FEM model assumes that the nanorod is forced into bacteria by a large external force leading to large deformation around the tip. Reproduced with permission under a Creative Commons License from ref [75]. © 2021 The Authors. (e) TEM images of deformation, and puncture of the cell wall of (i) gram-positive, (ii) gram-negative bacteria. Reproduced with permission under a Creative Commons License from ref [75]. © 2021 The Authors. (f) Schematic showing the tension imposed on the bacterial cell membrane by a specific nanopattern increases the degrees of membrane stretching. At an increased aspect ratio, an additional surface parameter, pillar bending enhances cell rupturing. (g-i) SEM micrographs of bacteria ruptured by flexible nanopillars on (g) dragonfly wing nanotopography, and (h) gram-negative rod-shaped bacteria or gram-positive cocci bacteria attached on flexible silicon nanopillars. (j) Another mechanism of mechanical damage is due to shear flow that brings bacteria to the pillars. Reprinted with permission from under a Creative Commons License from ref [76]. Copyright © 2023, The Author(s).

*Stretching and deformation* of the bacterial cell wall are actions associated with cicada-like nanopatterns (densely packed, high-aspect-ratio nanopillars that are not sharp enough to pierce the cell wall). The morphological changes forced upon the cell by physical conformations associated with surface adhesion lead to cell death[21]. The interaction between microbial-surface adhesion, cell elasticity, membrane rupture forces, and cell lysis at the microbial-nanostructure interface during the adsorption processes are all factors known to influence the mechano-bactericidal action of nanostructured surfaces. Several groups have observed that microbial morphology likely plays a role in the degree of antimicrobial activity on nanostructured surfaces[77,78]. In general, single-cell force spectroscopy studies utilizing AFM to probe the bacterial-nanointerface have shown that cells with a lower elastic modulus, and greater work of adhesion exhibited a reduced force required for membrane rupture and ultimately resulted in higher antimicrobial activity on those surfaces[79]. Studies of bacteria on smooth surfaces have demonstrated that a force of 20 nN is required to exert lethal damage to the cell membrane of *Escherichia coli*[80]. Generally, gram-negative rod-shaped bacteria are less rigid than gram-positive cocci bacteria, as indicated by indirect measurement of their Young's modulus (E values)[76]. Thus, the membrane of gram-negative bacteria deforms more easily under the same pressure. Cells that rupture more readily, and adhere more strongly to smooth surfaces, have a greater degree of cell lysis at the nanostructured interface. However, nanotopographical features reduce bacterial adhesion strength by reducing the available surface area[81]. AFM studies further revealed that bacterial adhesion strength was greater for more densely packed nanostructures and that cells with higher elasticity exhibited significantly stronger adhesion. EPS-producing strains demonstrated increased adhesion[81].

These phenomena may be partly explained by gram-negative bacteria possessing a thinner outer layer than gram-positive bacteria that have a 5 to 10 times thicker and more robust peptidoglycan layer that likely provides better protection against disruptive forces at the cell-nanostructure interface. However, a rod shape provides a larger surface contact area compared to the cocci shape, increasing the number of contact points with the nanostructured surfaces. This results in more significant cell deformation and higher levels of cell lysis. Indeed, finite element models have shown that Gram-positive bacteria consistently require a 6-fold larger force to yield/puncture than the Gram-negative bacteria[82]. It was recently discovered that mechanical injury from certain nanopatterned geometries was insufficient to immediately kill the bacteria owing to the survival of the inner plasma membrane[73]. Instead, such sublethal mechanical injury leads to apoptosis in the affected bacteria. In addition, when mechanical stress is removed, the self-accumulated reactive oxygen species (ROS) induce post-stress in damaged cells in a non-stress environment, revealing that mechano-bactericidal actions have sustained physiological effects on the bacterium[73].

Building on the biophysical model proposed by Pogodin *et al*., in 2013, several recent works have utilized the finite element method (FEM) to quantify the stress administered to the bacterial membrane as it interacts with nanopatterned topographies and, as such, determine the nanopattern geometries that lead to enhanced stretching and deformation of bacterial cell membranes, resulting in ultimate mechano-bactericidal efficacy[17, 74, 83-87]. In these models, the Von Mises failure criterion, which relates the calculated stress to the yield strength measured in a uniaxial tensile test, is adopted to assess the yield criteria of the bacterial membrane. In one example of a typical membrane model (4 nm thickness) of a gram-negative bacteria interacting with a nanopillar of 50 nm tip diameter, significant stress arises at the contact point (the pillar apex). The simulation reveals that the force needed to deform bacteria adhering to the nanopillar is much smaller (1.2 nN) compared to the force required on a flat surface (25.2 nN)[82]. Simulations of a gram-positive (20 nm) cell wall show that the *S. aureus* wall symmetrically bends around the nanocone, yielding an indentation depth of around 60 nm that creates a Von Mises stress value of around 13 MPa[17]. The biophysical model initially proposed by Pogodin *et al*. detailed that the adhesion of bacteria to nanopillars causes significant stretching of the cell wall, which can result in the cell wall rupturing between the points of adhesion, rather than localized puncturing. As a natural extension of this original analytical model, Alameda *et al*.[17] have performed detailed FEM simulations demonstrating that the maximum stress is observed in the parts where the cell wall is significantly more bent, that is, close to the detaching edges between the bacterium and nanocone, rather than at the nanocone tip. Islam *et al*. simulated a gram-negative bacterium adhering to a nanopillar array. The highest von Mises stress on the cell membrane around the contact points was estimated to be 6.97 MPa, which lies between the cell wall critical stress (5 MPa) and the tensile strength (13 MPa), thus leading to the creeping deformation of the cell wall. A higher strain was noticed at the local regions around this contact line and the nearby suspended regions of the cell rather than at the contact point itself [85]. Cui et al. found that the locations of maximum stress/strain occurred at the "three-phase contact line" (the contact line between the liquid-cell-nanopillar phases), neither at the cell suspension region between nanopillars nor at the cell-nanopillars contact region [74], which is in agreement with Valiei *et al*, and Zhao *et al*. [82, 88, 89]. However, some FEM models developed to investigate the interaction between bacteria and bactericidal nanopatterns incorrectly assume the application of external forces, such as gravity and the weight of the bacteria, or capillary forces. The FEM developed by Velic *et al*., describe a two-step process with time-dependent interactions. In their FEM simulations, external pressure is applied, squashing the bacteria on the pillars with maximum stress at the tips for any geometry (Figure 3b)[83]. Liu *et al*., postulate simultaneous piercing by all pillars for any configuration and deduce that if bacteria are pierced, it should be external force or pressure from above to push bacterium into the pillars[90]. In another example, an artificial neural network was created to analyze the response of three common bacterial species: *E. coli*, *Pseudomonas aeruginosa*, and *Staphylococcus aureus* to various geometrical features. Geometrical features that effectively target all three bacterial species were determined and were then used to create a series of finite element models. These models simulated the physical interactions between the bacteria and the nano-patterns, leading to bacterial inactivation; however, the model considered external forces such as gravity (the weight of the media, and of the bacteria itself) and did not consider adhesion forces. It should be noted that bacteria are not subjected to gravitational forces due to their negligible weight. Nevertheless, the authors determined bacterium's cell wall is stretched between the adjacent peaks of the nanopillars. The suspension of the bacteria causes further strains in the wall that increase to eventually reach the tolerable limit value for the bacteria[86]. Nevertheless, most theoretical studies have confirmed that membrane stretching and deformation (the impact of geometry of the nanopattern is discussed below) can lead to membrane rupture and this was also demonstrated in various experimental studies. Still, the forces required to rupture the membrane and localization of maximum stress incurred are under debate.

***Direct contact killing through piercing*** is associated with sharp nanopillars or nanospikes (tip diameter < 5 nm) that create stress points on the bacterial membrane upon contact[91]. These nano-feature-induced stresses lead to punctures or tears in the membrane from initial interaction with a sharp or uneven surface, eliciting cell lysis. In this regime of sharp nanofeatures, the bacterial response to adhesion involves a localized increase in stress on the tips of the nanopillars [74, 82, 83, 88]. For example, when the bacterial adhesion to a sharp-nanopillar array (of 16 nanopillars) was modelled, the stress upon the deformation remained focused on the contact points. This transition from stretching to piercing regime happens when the tip diameter becomes comparable wit the characteristic thickness of the lipid bilayer, as was discussed in a model of piercing of a lipid bilayer by a carbon nanotube, inserted perpendicularly[91]. Indeed, our recent transmission electron microscopy investigation of the cell-nanopillar interface following bacterial interaction with silicon nanopillar topography revealed the formation of holes in the bacterial membrane commensurate with the size of the nanopillars[30]. In another example, based on finite-element simulation, it was postulated that flat-top nanorod delivers the maximum stress of 5.6 MPa on a bacterial cell wall, lower than its ultimate-tensile-strength (13 MPa); while such sharp-tipped (3 nm) nanorods delivers a maximum stress of 20.1 MPa to puncture the cell envelope [92]. At 30 min incubation, both sharp-tipped and flat-tipped $Al_2O_3$ nanopillars penetrate the adhered bacteria at a depth of ~64 nm and 26 nm, respectively, as identified by analysis of TEM micrographs. It is suggested that the nanorods deform the cell envelope when the nanopillar tip has an angle θ larger than 138°, and penetrate the cell envelope when the tip θ is smaller than 138°[92].

*Induction of shear forces* is relevant in dynamic environments, such as the flow of fluid over surfaces, where shear flow generated by fluid motion can become sufficiently intense to cause damage to the bacterial cells. Surfaces that promote turbulent flow can cause bacteria to be physically swept away or subjected to high-stress conditions, resulting in mechano-bactericidal action[76, 93]. Carbon-coated, $Cu(OH)_2$ nanowires with tip diameter ~200 nm and lengths up to 5 μm that were grown on a copper foam substrate were demonstrated to efficiently inactivate bacteria even under mild water flow, such as that from a storage tank. The flow rate in the main flow was 2.7 mL min$^{-1}$. The bacterial cell lysis was suggested to be due to the strong shear flow between the nanotip surface and the bacterial cell envelope tearing the cell envelope rather than from collisions[76]. The authors modelled the stress distribution profile of a cell membrane based on its interaction with the nanotips. During collision without flow, the maximum stresses exerted on bacteria encountering the nanotips were calculated to be between 2-4 $10^{-4}$ MPa, which is lower than the critical value of 0.05 MPa calculated as required to effectively cause cell lysis. Under fluid flow, the maximum outward stresses were between 5-6 MPa, higher than the minimum stress required to rupture bacteria encountering the nanotips. However, the finite element method was used to determine the relative stress exerted on the bacterial cell membrane (modelled as a simple lipid bilayer) considering the adhesion of the cell membrane to a nanotip an order of magnitude larger than what has previously been modelled to be able to pierce a lipid bilayer (< 2 nm). Overall, the bactericidal performance of nanostructured surfaces is influenced by the scale, shape, material, and elasticity of the nano-features on the surface, the type of bacteria, and environmental conditions. Understanding the underlying mechanism of nanopatterns is imperative for designing surfaces that can be incorporated into specific applications.

*Flexibility/aspect ratio.* The flexibility and aspect ratio of the nanopatterns play a pivotal role in enhancing the mechano-bactericidal properties of engineered surfaces. The aspect ratio, defined as the relationship between the height and diameter of the nanostructures, directly influences the mechanical interactions between the bacterial cell membrane and the nanopatterned surface. High-aspect-ratio structures such as elongated nanopillars or nanowires exhibit greater flexibility, allowing them to bend under minimal force (10 nN – 200 nN) while maintaining contact with the bacterial membrane (Figure 3). Enhanced killing efficiency has been reported for these patterns which is attributed to the release of stored elastic energy in the pillars upon deflection[94, 95]. Pirouz *et al.*,[96] recently postulated that bacterial adhesion alone is insufficient to cause deflection of the nanopillars they encounter, regardless of their respective material Young's modulus. Instead, they propose that additional forces must be at play including the production of EPS, and cell migration. However, previous research by our group using EPS-deficient cell mutants has demonstrated that EPS does not influence the mechano-bactericidal efficacy of rigid nanopillars[69], and works reporting the deflection of nanopillars in response to bacterial adhesion, leading to enhanced mechano-bactericidal efficacy also show pillar bending in response to the adhesion of non-motile bacteria[95].

Pillar flexibility is crucial for exerting continuous mechanical stress on the cell, leading to membrane deformation, disruption, and ultimately cell death. Research has shown that nanostructures with higher aspect ratios are more effective at penetrating and breaching the bacterial membrane because their flexibility allows for a more dynamic interaction with the cell[104]. Moreover, the ability of these flexible, high-aspect-ratio nanostructures to adapt their shape in response to the bacterial cell morphology enhances the bactericidal efficacy of the surface by increasing the points of contact and, thus, the mechanical stress imposed on the cell. We have demonstrated that silicon nanopillar arrays with an assumed Young's modulus of $E = 66.5$ GPa. Imaging of the bacterial-nanointerface allowed for direct estimation of the deflection of each pillar to calculate the mean absolute forces and mechanical energy of the pillar tips resulting from lateral deflection. There was an increase in mechanical energy and elastic forces for nanopillar arrays lengths 360 and 420 nm, which led to additional stress being placed on the bacterial cell membrane. This correlated with enhanced mortality observed for *P. aeruginosa* and *S. aureus* attached to taller pillar arrays over shorter (200 nm) pillar arrays[95].

With respect to polymeric nanopatterned surfaces, increased pillar stiffness has been correlated to increased bactericidal efficacy of the nanoarray. For example, ultraviolet nanoimprint lithography (UV-NIL) was used to fabricate flexible nanopillar arrays of controlled dimensions and varying elasticities from 208 MPa to 4 GPa[97]. Experimentally, thin (≤100 nm) and stiff (≥1.3 GPa) nanopillars compromised bacterial viability, while thick (200 nm) and flexible (<1 GPa) nanopillars did not increase the proportion of dead cells compared to the control. Increasing the pillar stiffness and reducing the pillar diameter were found to result in greater reduction in bacterial viability[97]. It has been routinely noted that the nanopillars bend at the external perimeter of the cells whereas the pillars underneath the center of the cells remain perpendicular and independent[95]. In another work, a nanopillar array resembling moth eyes were fabricated on PMMA and hPDMS. There was a 4-fold increase in cell death of *S. aureus* when cultured on patterned PMMA substrata (higher stiffness, E~GPa) compared to that on moth-eye patterned hPDMS substrata (lower stiffness, E~MPa)[17]. Biomimetic nanotopographies with varying aspect ratios were created using maskless dry etching of PET. It was observed that structures with both high and low aspect ratios effectively damaged the membranes of Gram-negative bacteria. However, these structures did not inactivate Gram-positive bacteria. Additionally, the clustering of the soft, flexible tall (500 nm) nanopillars led to cooperative stiffening, a phenomenon confirmed through nanomechanical analysis and supported by finite element simulations[98].

Therefore, designing nanopatterns with optimal aspect ratios and inherent flexibility targets bacterial cells more effectively and introduces a mechanically adaptive component to bactericidal surfaces, paving the way for developing more efficient and broadly applicable antimicrobial materials.

## 3.2 Correlation of Bactericidal Efficacy with Surface Topography and Geometry of the Nanopatterns

The simple elastic model of the mechano-bactericidal action of the nanopillar topographies considers four key geometric parameters of the surface features: spacing, tip diameter, base diameter, and height (Figures 2-3). In most cases, these selected parameters were sufficient to describe the mechanical killing of bacteria by *nanopillar* topographies adequately (Supplementary Table 1S). However, not all nanopatterns can supply the necessary mechanical forces to kill all the attaching bacterial cells. Furthermore, recently it was reported by several groups that specific nanopatterns, distinct in their geometry, engage in various direct physical interactions with bacterial cells, demonstrating a range of lethal effects[78, 94, 95, 99-101]. Velic *et al*. explored how geometric features of nano-patterned surfaces—such as peak sharpness, height, width, aspect ratio, and spacing—affect their mechano-bactericidal properties[84]. The finite element method was used to analyze cell-nanostructure interactions with *E. coli*. The numerical solutions obtained were then verified with artificial neural network (ANN) methods. An increase in peak sharpness, aspect ratio, and spacing led to higher maximum deformation, stress, and strain on the *E. coli* cells[102].

To fully comprehend how different patterns confer antibacterial activity, here we examined how each element influenced the interaction between the surface and bacterial cells. Based on our systematic literature analysis of mechano-bactericidal patterns, the pillar density and height are the key surface parameters influencing bactericidal efficacy.

*Size.* Size and shape are crucial factors in determining the effectiveness of nanostructures in rupturing bacterial cells[14, 103]. For instance, nanopillars or nanospikes must have suitable dimensions to interact with bacterial cell membranes physically. If they are too large, they may not exert sufficient localized pressure to penetrate the cell membrane. If they are too small, they may be unable to pierce the robust outer structure effectively. Cui *et al*. determined that a critical nanopillar height of approximately 200 nm was required to kill *E. coli* cells upon contact of ∼200 nm[74]. At a fixed pitch and pillar diameter, taller nylon nanopillars (120–220 nm) showed increased bactericidal and antifouling efficacy toward both *S. aureus* and *P. aeruginosa* compared to short nanopillars[104].

*Shape.* Shapes play a considerable role in the bactericidal effectiveness of nanostructures. Sharp tips or edges are more effective at inducing stress points[92, 105], which aid in breaching bacterial cell walls, compared to blunted or rounded structures that may lack the same effect[100]. For example, polycarbonate surface nanopillar patterns with smaller pillar cap diameters (sharper tips) were more effective in killing *E. coli* cells than blunt pillars[106]. Finite element simulations performed in COMSOL confirmed that due to their smaller tip area, the stress induced by thin polymer pillars is larger than that of thicker pillars[97]. In a comparison of cyclo-olefin polymeric nanocones *versus* nanopillars, nanocones, which have a higher aspect ratio and height than nanopillar structures, demonstrated less bactericidal efficacy than nanopillar structures, especially those with a height of 300 nm, and periodicity of 200 nm. Nanocones exhibited a much-reduced elastic modulus, compared to the nanopillars, and a comparatively lessened ability to inactivate contacting bacteria, despite the sharper tip. Finite element simulations showed that the amount of bending deformation in nanopillar structures is positively correlated with their bactericidal effectiveness, while the opposite holds true for nanocone structures [107]. The shear stress applied to the resin nanostructures during contact with *E. coli* cells caused the nanostructures to bend and deform. Increasing the local elastic modulus of nanopillar structures increases the mechanical bactericidal performance of their topology. Conversely, for nanocone structures, the situation was reversed. By contrast, in a recent study on polymeric cicada-like nanocone arrays, Zhao *et al.* have shown that the bactericidal efficiency against *E. coli* increased by 5.5%~31%, when exposed to nanocone arrays with sharp tips and larger interspaces[89, 107]. A finite element model revealed that varying nanostructures impact bacterial adhesion by modifying the contact area with bacteria and causing stress and deformation on the cell wall. Nanostructures with smaller tip diameters reduced the area of contact with bacteria, which diminished the strength of bacterial adhesion [108].

*Spacing/periodicity.* The spatial distribution and arrangement of the nanostructures also play a significant role. The surface must be designed so the bacterial cells cannot fully conform to the nanostructures without encountering mechanical stress points. Optimal spacing ensures that the cells are stretched across multiple nanopillars, leading to multiple points of membrane tension and eventual rupture[74, 85, 89]. Most mechano-bactericidal nanopatterns reported in the literature exhibit an interspacing of less than 300 nm[14]. Several theoretical and experimental reports have concluded that nanopillar patterns with short features and increased pillar density (*e.g.*, heights of 200 nm or less and spacings of 100 nm or less) exhibit greater degrees of mechano-bactericidal efficacy than patterns containing more widely spaced, larger features (*e.g.*, heights greater than 300 nm and spacings greater than 300 nm[106, 109]). Dickson *et al.* found that polymethylmethacrylate surfaces with closely packed nano protrusions exhibited greater bactericidal activity toward *E. coli* than nanopatterns with larger spacing; the minimum threshold for optimal nanopillar spacing was between 130 and 380 nm[122]. This trend was confirmed by Hazell *et al.*, using PET nanocone arrays. In our recent study, very dense or less dense features of acrylic nanopillar patterns were confirmed to result in greater bactericidal efficacy than those with intermediate spacing. Specifically, acrylic films with a nanopillar pitch p = 60 nm showed the highest rates of antibacterial activity (approximately 90% of attached cells were non-viable) toward *P. aeruginosa*. Increasing the nanopillar pitch led to a gradual decrease in the antibacterial performance (bactericidal and anti-biofouling properties) of the respective surfaces. In turn, samples with either 60 nm or 200 nm pitch were the most effective at inactivating attached *S. aureus* bacteria[101]. The same trend was observed for both bacterial species at pillar aspect ratios of 1 and 4, confirming that the aspect ratio did not significantly influence bactericidal performance. Simulations comparing the bacterial deformation on an array of 16 nanopillars, with the geometry constructed in two arrangements with differing array densities, a compact array with a centre-to-centre interpillar spacing of S = 60 nm and a widely spaced array with S = 100 nm found that the more compact nanoarray required lower amounts of force (3.3 nN) to result in membrane yield than the widely spaced nanopillars (6.9 nN)[82]. This trend is further corroborated by Velic *et al.*, model using tightly packed nanopillars of a small tip radius, which can simultaneously elicit high areal stresses and high contact pressures, as conveyed by the combined von Mises stress[84]. Older models of the bactericidal mechanism of nanopatterned surfaces[110] implemented a similar energy functional—involving Helfrich curvature-elasticity and thermodynamic adhesion energy—to study the interaction of bacteria on spherically-capped cylindrical nanopillars. They argued that nanopatterns with larger nanopillar radii and higher nanopillar density (smaller centre spacing), would yield the greatest killing efficiency by enhancing areal strain. Comparing nanopillar radii $0 \leq r \leq 50$ nm, and spacing $100 \leq s \leq 250$ nm, maximum areal strain on the envelope was induced by combining the largest radius (*i.e.*, r = 50 nm) with the smallest centre spacing (*i.e.*, s = 100 nm). However, in these studies, out-of-plane effects were ignored. Velic *et al.*, confirmed what experimental studies have shown, that too densely packed pillars will actually support the bacterial membrane and result in a loss of bactericidal efficacy. Therefore, an excessively high density of structures can provide enough adhesion points on the nanostructures to prevent stretching between the structures, thereby reducing the stress experienced by adhered cells. As a result, this can diminish the bactericidal effectiveness of the nano-structured surfaces.

## 4. ANTIFUNGAL SURFACES

### 4.1 Antifouling Surfaces

Nanostructured surfaces can be fabricated to disrupt fungal cells mechanically through cell–wall interactions. Simultaneously, nanostructured surfaces designed to be superhydrophobic due to their surface architecture can serve a dual purpose by exhibiting self-cleaning abilities[18, 58, 63, 66, 111]. Research in this domain continues to evolve, potentially bringing forth advanced surfaces that can be integrated into medical devices, food packaging, and agricultural products to safeguard against fungal threats.

Recently, the surfaces of dragonfly wings were reported to repel fungal spores (also referred to as conidia), preventing fungal colonization (Figure 4)[27]. This remarkable property was found to be due to air bubbles becoming trapped within the nanoscale pattern of the wing topography. The entrapped air layer has several purposes: first, the lack of a water film makes it challenging for fungal spores to adhere to the surface and germinate. For many fungi, surface moisture is a prerequisite for initiating the growth cycle; therefore, the absence of this critical moisture layer can inhibit fungal development. Secondly, surfaces that utilize this mechanism likely minimize the risk of biofilm formation especially of fungal conidia, which are significantly larger (up to 10 times) than bacteria, to protect against environmental stress and antimicrobial treatments.

Nanotexturing can be used to design hydrophobic coatings by including micro- and nano-scale roughness, which reduces surface energy and water retention[112]. When applied to medical device design, packaging, or other materials, these principles can considerably decrease the risk of fungal adhesion, proliferation, and resultant complications[113]. The manipulation of surface wettability has shown promise as a non-toxic and long-lasting method for resisting fungal contamination[47, 112]. There have been several excellent reviews of surface modification techniques for anti-wetting systems to prevent biofouling[60, 63, 66, 111, 114]. Therefore, the next sub-section focuses on surface nanotopographical modifications that are fungicidal.

### 4.2 Physical Rupture of Fungi by Nanostructured Surfaces

Unlike the mechano-bactericidal mechanisms of nanostructured surfaces, the multiple mechanisms of action of fungicidal nanostructured surfaces have not yet been defined. Nanostructured surfaces designed for fungicidal applications must account for the physical robustness of the fungi. One of the primary physical mechanisms by which nanostructured surfaces exhibit antifungal properties is the disruption of the fungal cell envelope. However, fungicidal nanopatterned surfaces may be required to deliver stronger mechanical forces to disrupt the cell walls[115]. Unlike bacterial cells, fungal cells and spores are eukaryotes with membrane-bound organelles and possess a more complex and robust cell wall structure composed of multiple layers, including chitin[116]. The chitinous cell wall contributes to the cell wall rigidity and is a stratified structure consisting of chitinous microfibrils embedded in a matrix of small polysaccharides, proteins, and lipids. Internal to the cell wall lies a plasma membrane and the cell integrity is further maintained by microtubules composed of tubulin. Yeasts are unicellular fungi that reproduce by budding, whereas mold forms multicellular hyphae [117].

As such, larger or more sharply defined nanostructures may be necessary to puncture the cell walls of fungi (0.1 to 1 µm), compared to those effective against bacteria[118]. The spacing between features also plays a crucial role because proper alignment can hinder the ability of the hyphae to anchor and grow across the surface area. However, like bacteria, yeasts have shown susceptibility to nanostructure-induced rupture on the nanotopography of cicada wings (Figure 4, Supplementary Table 1S)). *C. albicans* cells were found to be ruptured and killed following attachment to the nanostructured wing surface of cicada *Neotibicen tibicen*. The wing topography of *N. tibicen* cicada is different to that of *P. claripennis* cicada and features an array of nanoscale cones each measuring 200 nm in height and width, a 30 nm tip diameter, and periodicity of 200 nm [24].

Recently, we demonstrated that silicon nanospikes 800 nm in height, with a tip-to-tip separation of 200 nm, were sufficient to induce the physical rupture of *Aspergillus brasiliensis* conidia (Figure 4). [118]. The resulting loss of critical intracellular contents compromises the ability of the fungus conidia to maintain its internal physiology, ultimately triggering programmed cell death[119]. By contrast, silicon micropillar arrays with 3.5 µm height spikes and separation of 3.7 µm exhibited an increase in yeast cell *C. albicans* adhesion and proliferation compared to the flat control samples. The large separation between the pillars meant that yeast cells could align themselves to maximise surface contact. Hyphae production was observed and they seemed to use tip contact to support continued linear growth[115]. When the interpillar spacing is less than the length of the body of the fungi, cell viability has been shown to decrease [24, 118, 120]. For example, highly ordered $TiO_2$ nanotubes of 80 nm diameter and 400 nm height and interpillar spacing 20 nm, reduced *C. albicans* adhesion and viability and cells did not form pseudo hyphae. Similarly, we recently showed that Ti surfaces with varying degrees of nano roughness affected the colonization of *C. albicans*. Smooth Ti surfaces (with an average surface roughness of 25.7 ± 8.5 nm) supported similar levels of viable *C. albicans* as the nanorough Ti surfaces (with an RMS roughness of 484.0 ± 15.6 nm) after 24 h. However, the polished Ti surfaces led to a pseudo hyphal form of the cells, while the nanorough Ti surfaces resulted in a yeast-like ovoid morphology [113]. Therefore, the adhesion of yeast to nanostructured surfaces not only results in the loss of cell viability, through cell wall-nanostructure interactions, but also alters the cell phenotype resulting in reduced pathogenicity[24, 121].

It is still unclear how an external mechanical stressor inhibits hyphal morphogenesis. Upon attachment to nanostructured surfaces, *C. albicans* cells exhibited a dramatic change in membrane potential demonstrating that the cells experience the application of significant external stress when adhering to nanostructured surfaces [23]. Indeed, several groups have also reported the perturbation/indentation of the cell wall at the cell-nanostructure biointerface[118-120]. A transcriptomics study of *C. albicans* cultured on cicada wing nanotopography determined changes in the expression of genes associated with metabolism, biofilm formation, plasma membrane component biosynthesis, and DNA damage response after 2 h of exposure[23]. Similarly, we recently showed the deterrence of *C. albicans* biofilm formation on Ti nanopillared surfaces. Proteomics analysis revealed down-regulated proteins that are involved in the development of filamentous/hyphae biofilm phenotypes of yeast cells. Furthermore, the disruption of the cell wall induced by nanostructured surfaces leads to the production of proteins involved in rebuilding the cell membrane[23, 119]. In addition, the expression of several proteins, including metacaspases and those involved in the mRNA process, chromatin remodeling, and nucleosome assembly, provides compelling evidence that *C. albicans* cells undergo apoptosis following interaction with nanostructured topography even when the mechanical injury was insufficient to immediately kill the cells[119].

In a recent work, it was shown that *C. albicans* and *C. neoformans* cells exhibited similar adhesion to both unmodified and nanostructured titanium surfaces. However, *C. neoformans* cells are more susceptible to rupture, revealing up to ~80% cell lysis at the nanospike titanium interface. Indeed, based on AFM data it was revealed that *C. neorformans* exhibited a lower rupture force (Fr) value, and deformed more readily, as shown by lower Young's modulus (E) values, compared to *C. albicans* cells. This greater elasticity may be attributed to higher levels of chitin and chitosan in the cell wall of *C. neoformans*. Conversely, *C. albicans* was more resistant to the antimicrobial action of the nano-structured titanium for both nanowires and nanospikes surface architecture. For both bacteria and fungi, increased cell death on nanostructured surfaces is correlated with a lower rupture force and higher cell wall elasticity[79].

While numerous works describe the fungicidal action of nanostructured surfaces toward yeasts, there are limited studies investigating the physical rupture of environmental, filamentous fungi on nanostructured surfaces. Indeed, we recently reviewed the impact of multiscale surface topography on *C. albicans* biofilm formation [48].

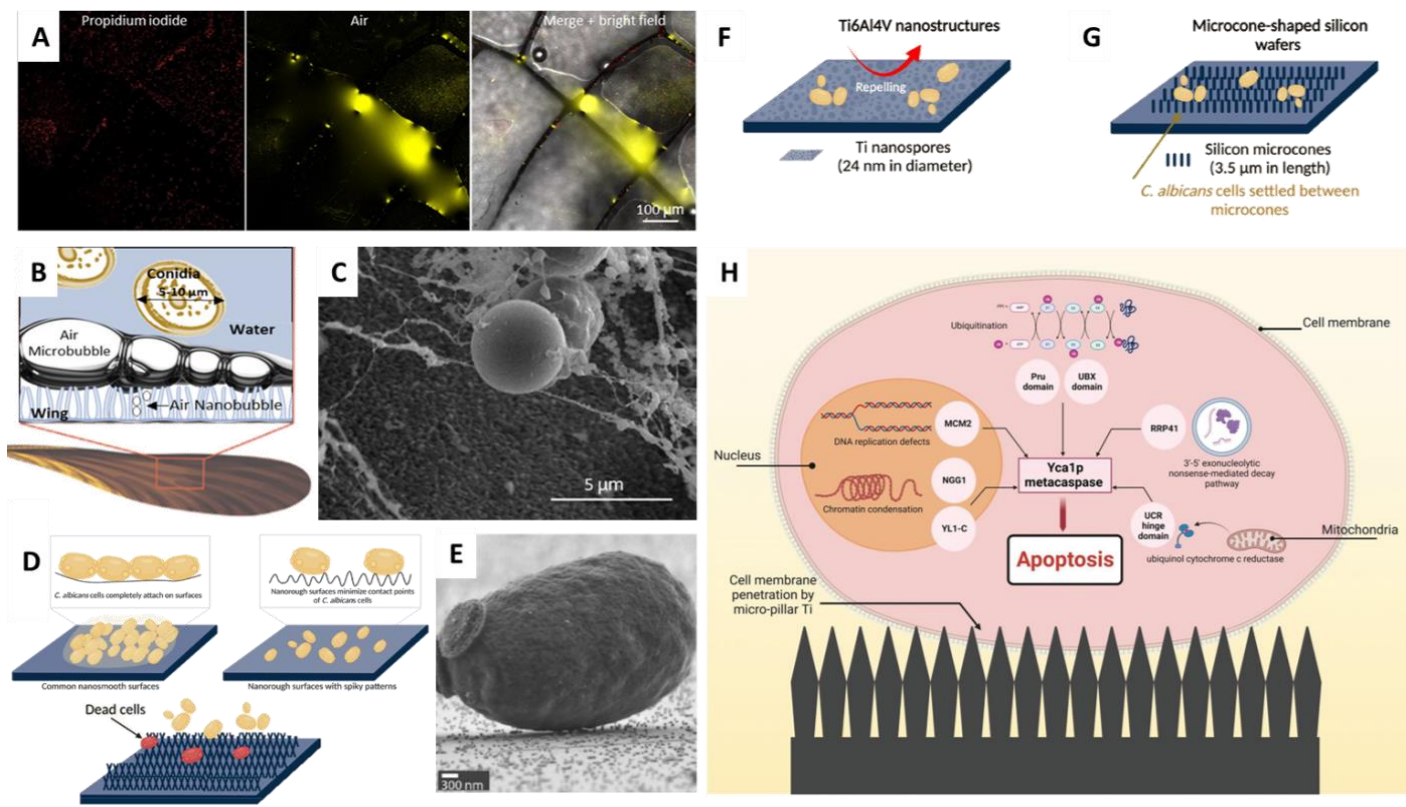

**Figure 4. Antifungal nanostructured surfaces**. (a) Fluorescence micrographs showing bacteria (red, propidium iodide) crossing the air-water interface (yellow) to attach on the superhydrophobic wing surface. (b) schematic representation of fungal conidia repelled from attaching on the wing surface by an air cushion. (c) SEM micrograph of fungal conidia of *Aspergillus brasiliensis* trapped in the water layer above the wing surface. Images A-C reprinted with permission from ref [27]. Copyright © 2021 Elsevier Inc. (d) Schematic of *Candida albicans* attachment on smooth (top-left), nano-rough (top-right) or high-aspect-ratio nanostructured surfaces (bottom). (e) SEM micrograph of fungal conidia sitting on top of nanoparticle coatings on PDMS. Antifungal micro-nanostructured metallic surfaces: schematic of antifungal (f) nanoporous titanium alloy Ti6Al4V and (g) micro-cone topography. Panels D-G reprinted with permission under Creative Commons License from ref [48]. Copyright © 2024 The Author(s). (h) Micro-nanostructured titanium surfaces trigger apoptosis in attaching *C. albicans* cells. Images reprinted with permission under Creative Commons License from ref [119]. Copyright © 2023 The Author(s).

# 5. VIRUCIDAL MECHANISMS

## 5.1 The Nature of Virucidal Mechanisms Originating from Nanostructured Surfaces

Fomite transmission, the secondary transmission of viruses through contact with contaminated surfaces, is a significant route for infections, especially in high-density residential, healthcare, and public settings. Few groups have demonstrated that various viruses, including influenza, norovirus, and human coronaviruses like SARS-CoV-2, can remain viable on surfaces for extended periods, ranging from hours to days, under ambient conditions[35, 122]. The survivability of these pathogens on surfaces underscores the critical need for effective disinfection protocols and developing surfaces with inherent antiviral properties to mitigate the risk of fomite transmission. However, the complexity of virus-surface interactions, influenced by surface material, environmental conditions, and specific virus characteristics, poses challenges in formulating universally effective strategies to combat surface-mediated transmission.

Nanostructured surfaces can inactivate and mechanically destroy viruses, which are considerably smaller and structurally different from bacteria and fungi[123]. Leveraging nanostructures to incapacitate viruses by physical interactions is challenging, given viral particles' compact and sturdy nature. We recently proposed that the height and spacing of nanopatterns should be smaller than the virus size to help penetrate the viral envelope and/or reduce the contact area, thus reducing viral adhesion[124]. The physical rupture of viral particles on nanostructured surfaces often relies on the distortion and subsequent rupture of the viral capsid, a protein shell encasing and protecting the genome. Nanoscale surfaces can induce sufficient mechanical stress upon contact with the viral particles, leading to physical disruption. This damage can cause the release of viral genetic material, rendering the virus noninfective. Nanoscale roughness, sharp edges, and spike-like features are particularly effective against viral particles because of their high surface-area-to-volume ratio and increased likelihood of contact (Supplementary Table 1S)[30, 124, 125].

The ability of nanostructured surfaces to inactivate or disrupt viruses on specific surfaces is associated with the mechanical force of nanopillar features, shear flow around sharp pillars, and entrapped air layers[76, 126].

*The mechanical force of nanopillar arrays*. These arrays comprise regularly spaced nanoscale protrusions that could interact with viruses upon contact. The mechanism of action is primarily physical: nanopillars apply mechanical stress to the viral particles or pierce as they come into contact with the surface, potentially causing deformation or rupture of the viral capsid or envelope (Figure 5). The effectiveness of nanopillar arrays as virucidal surfaces depends on their size, shape, spacing, and height, which must be optimized relative to the size and morphology of the target viral particles. For instance, coronaviruses are decorated with spike proteins that protrude from their envelope, and a surface with appropriately spaced nanopillars may interlock these spikes, causing mechanical stress that deactivates the virus. The photocatalytic properties of nanoarray surfaces or those that produce reactive oxygen species may amplify virucidal activity when engineered into nanopillar arrays[127]. Hasan *et al*. investigated the effectiveness of nanostructured surfaces toward inactivating respiratory syncytial virus (RSV), and rhinovirus (RV). Hydrothermal etching of aluminium was used to create a random distribution of sharp (23 nm) nanowires. The recovery of infectious RSV from the etched surface was much lower than the control surface. The nanostructured surface proved more effective against the non-enveloped RV than the enveloped RSV virus[128]. In a related study, the same researchers found that exposure to the nano-patterned surface for 6 h resulted in complete inactivation of severe acute respiratory syndrome coronavirus 2 (SARS-CoV-2) [129]. However, the exact mechanism of inactivation remains unverified and is speculative.

One reported example of a virucidal nanostructured surface that has gained attention is black silicon, the first synthetic biomimetic analog of the nanoarchitecture of dragonfly wings[19]. This design exploits the natural high-aspect-ratio nanopillars found on wings, which have demonstrated remarkable efficacy in disrupting bacterial cells and inactivating viruses. The sharp tips of these nanopillars were in the range of tens to hundreds of nanometres (1-2 nm) in diameter, and the pillars were several hundred nanometers (300 nm) (Figure 5). The pillars were spaced at intervals (60 nm) comparable to the size of virus particles, typically ranging from tens to hundreds of nanometers in diameter. Such close spacing ensures that the viral particles coming into contact with the surface will likely interact with multiple pillars simultaneously. As the virus particles settle on the surface, they are subjected to mechanical stress at the points of contact with the nanopillar tips. These stresses cause deformation and rupture of the viral envelope, leading to virus inactivation[30]. The sharp features of nanopillars are crucial; they increase the likelihood of mechanical penetration through the envelope or capsid, particularly for enveloped viruses such as the influenza virus or SARS-CoV-2[30].

We recently showed that silicon (Si) surfaces with sharp nano spikes, with an approximate tip diameter of 2 nm, caused a 1.5 log reduction in the infectivity of human parainfluenza virus type 3 (hPIV-3) after 6 h. Finite element modelling of virus-nanospike interactions suggests that the virucidal effect is due to the sharp nanofeatures' ability to penetrate the viral envelope[30]. Similar to the mechano-bactericidal efficacy of nanostructured surfaces, the effectiveness of viral inactivation on nanostructured surfaces may vary depending on the specific surface nanopattern and the viral strain. For instance, nanostructured aluminum surfaces inactivated SARS-CoV-2 more rapidly than RSV and rhinovirus (RV)[128]. In contrast, Hasan *et al*, reported that nanostructured titanium surfaces achieved a 2.6 log reduction in RSV viral infectivity after just 5 h, likely due to a synergistic effect of photocatalytic effect of titanium surfaces via the generation of reactive oxygen species (ROS) [125].

Regarding the mechanical inactivation of viruses with nanopillars/nanospikes, we propose distinct virucidal mechanisms depending on whether the virus interacts with one, two, or four spikes. For a single nanospike with a tip diameter of 2 nm, or less, we suggest a piercing mechanism. According to COMSOL simulations, when spherical viruses interact with a single nanocone with a 1-nm tip diameter, overcoming the interaction energy barrier (∼100 kT) to pierce the viral envelope requires an external force. While the viral envelope is modelled as a tensionless elastic lipid bilayer, the hPIV-3 envelope is known to have a dense layer of glycoproteins and undergo changes in tension upon attachment to surfaces. Stiff membranes, which deform less, experience a higher concentration of stress at the tip apex and thus require less force to penetrate. The Young's modulus of a viral capsid is approximately 1–3 GPa. The sharp tip of the nanocone combined with the membrane's extreme stiffness facilitates the piercing process[30].

Furthermore, while surface interactions are assumed to be attractive and governed by van der Waals forces, protein amino acid groups in the membrane, such as −NH2, −NH3+, −COOH, and COO, drive adsorption through electrostatic interactions and hydrogen bonding, particularly at neutral pH, where viral particles are generally negatively charged. This enhances membrane tension. In the presence of a thin water film, viral particles can establish stronger adhesion through hydrogen bonding between OH groups at the surface and the virus surface proteins. AFM measurements showed that the adhesion force of MS2 coliphage to nanostructured Si surfaces is 3.3 nN, compared to 5.2 nN for smooth Si surfaces, indicating a stronger adhesion than required to rupture a Gram-negative bacterium[130, 131]. Additionally, the free energy of interaction for RSV, closely related to hPIV-3, with silica surfaces was estimated to be $5.5 \pm 0.4$ mJ m$^2$, indicating a net attractive interaction[132].

When a virus interacts with more than two nanospikes, COMSOL simulations suggested that the induced stress is distributed across the membrane area between the nanospike tips, rather than being concentrated at the tip apex. This results in a mechano-bactericidal effect similar to that observed with nanostructured surfaces, where membrane stretching between nanoprotrusions overcome the membrane's elasticity, leading to rupture.

***Shear flow around sharp pillars*** is an intriguing aspect of fluid dynamics that has implications for the designing and functioning of nanostructured surfaces with virucidal properties. Nanopillars interact with fluids and any viral particles contained within them that flow across their surface. The geometry of these pillars substantially affects the behavior of the fluid flow and can induce local shear forces with sufficient strength to impact the integrity and infectivity of viral particles, similar to bacteria[93, 133, 134]. When a fluid passes over a nanostructured surface, the velocity gradient between the fluid layer adjacent to the surface and the layers above it creates shear stress[93, 133]. This effect is more pronounced around the tips of the sharp pillars, where the fluid must navigate around the nanostructures. The shear forces generated in such scenarios can physically damage the viral particles in several ways. If the shear stress is sufficiently high, it can cause mechanical deformation of the viral capsid or envelope, especially if the viral particle is caught between the fluid flow and the nanopillar tip. These deformations can be significant enough to rupture the viral envelope or capsid structures, which are essential for the ability of the virus to infect host cells.

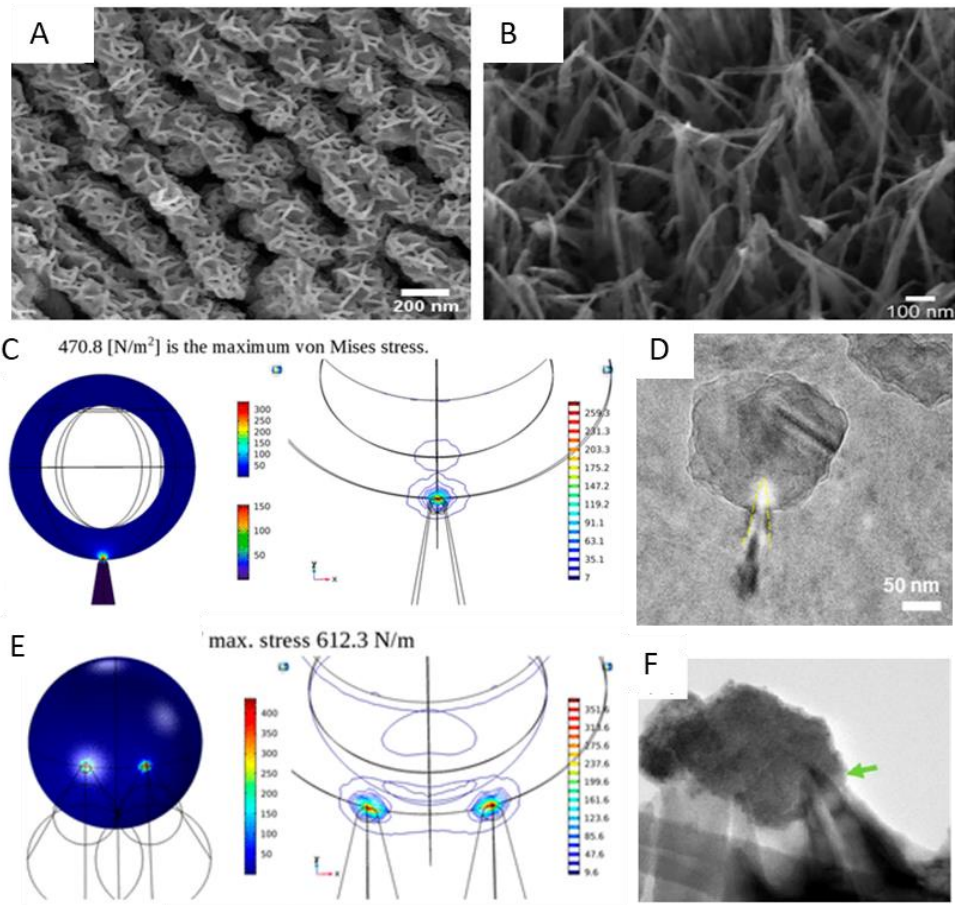

**Figure 5. Nanostructured topographies with reported antiviral activities.** (a) Chemically etched surfaces with nano-ridges fabricated using the wet etching of aluminium. Reprinted with permission from ref [128]. Copyright © 2020 American Chemical Society. (b) Nanostructured TiO$_2$ nanowire surface fabricated using hydrothermal treatment. Reprinted with permission from ref [125]. Copyright © 2023 Elsevier Ltd. (c) Finite element method model of mechanical disruption of human parainfluenza viruses on a single silicon nanospike and (d) corresponding and (d) TEM image of the virus at the tip of the spike. (e) finite element modelling of the interaction of the virus with multiple nanospikes and (f) corresponding TEM image of the virus with multiple spikes piercing the virus. Panels C-F reprinted under Commons Attribution Non-commercial License 4.0 (CC BY-NC) from ref [30] © 2024, The Author(s).

Induced shear flow can also remove surface proteins that many viruses rely on for cell attachment and entry. Moreover, the local increase in the shear rate near the sharp pillars can promote the dispersal of viral particles, preventing the formation of concentrated areas of viruses that could otherwise lead to higher risks of infection transmission. This dispersion is particularly relevant in scenarios where the fluid flows over the surface repeatedly or continuously, as in the case of catheters, filters, and ventilation systems[135].

The effectiveness of shear flow around sharp pillars as a virucidal mechanism depends on several factors, including fluid properties such as viscosity and flow rate, geometric arrangement, size and spacing of the pillars, and the size and morphology of the virus itself. For instance, enveloped viruses with flexible lipid bilayers may be more susceptible to shear-induced deformation than non-enveloped viruses with more robust capsids[123].

The design of these nanopillar arrays considers the conditions under which virucidal surfaces function. In high-flow environments, pillars must be sufficiently sturdy to withstand continuous fluid stress without degrading or detaching from the surface. Material selection is crucial for ensuring both the mechanical stability of the pillars and the antiviral efficacy of the interaction.

*Entrapped air layers* operate on the principle of creating a superhydrophobic surface, akin to the lotus leaf effect, in which microscopic and nanoscopic surface structures trap air in their interstices[58, 136]. When a virus-containing droplet approaches a surface, it does not adhere to or spread across it. Instead, it beads up and rolls off owing to the cushioning effect provided by the air layer. A surface that minimizes liquid retention can reduce the survival time of viruses, thereby demonstrating virucidal or virus-preventive capabilities. The transition from complete wetting in the Wenzel state to the air-entrapped Cassie–Baxter state on the nanopillar structures of cicada wings has been studied for self-cleaning surfaces[111]. Entrapped air layers limit the exposure of the surface to viral particles and decrease the potential for fomite transmission, which is an essential aspect of infection control, particularly in public and healthcare settings. Chatterjee et al. found that adjusting both the wettability and roughness of a surface, regardless of its geometry, can decrease the lifespan of residual thin wet films. Surfaces with taller and more densely packed nanopillars result in the fastest drying times and the most effective antiviral properties[124]. Nanoporous silicon was shown to significantly reduce both the number of viruses adhering and the strength of their adhesion. The nanohole array possessed diameters of 50 nm, depths of 22 nm, and pitch distances of 100 nm. They examined the adsorption of male-specific coliphage MS2 on both smooth and nanostructured surfaces using AFM imaging after 2 h of incubation with coliphage MS2. The nanostructured surface effectively reduced the adhesion of virus particles with most particles found between adjacent nanoholes. After 2 and 24 h of incubation, only 0.3% and 2% of the nanoholes contained coliphage MS2, respectively. This reduction in viral adhesion was attributed to the pseudo-Cassie–Baxter state of surface wettability, where air pockets in the nanoholes prevent coliphage MS2 viruses from attaching. Additionally, the small diameter (50 nm) and pitch distance (100 nm) of the nanoholes create limited space for viruses to land and aggregate. In contrast, on a PVC surface with randomly distributed nanopores (60 nm – 80 nm wide and 20 nm –30 nm deep), coliphage MS2 adhesion and retention were significantly higher, with virions found in every pore. The surface roughness of the nano-patterned silicon was 1.7 nm, compared to 10.8 nm for the porous PVC surface. The larger pore size, increased surface roughness, and overall heterogeneity of the PVC surface contributed to greater coliphage MS2 retention by increasing the contact surface area[131].

To integrate these specific surface properties into the design of virucidal materials, analyzing the conditions in which they will be used is crucial. For example, in high-touch areas, the mechanical durability of the nanopillar arrays may be a concern. For superhydrophobic surfaces, the presence of organic matter and cleaning chemicals may affect the longevity of the entrapped air layers. While nanopillar arrays and shear flow work predominantly through direct interaction and mechanical disturbance by viruses, entrapped air layers exhibit virus-preventive behavior by reducing surface contact and liquid retention.

# 7. CONCLUSION AND FUTURE WORKS

## 7.1 Insights Towards the Design of Nanostructured Surfaces with Enhanced Biocidal Efficacy

The investigation of nanostructured surfaces with bactericidal, antifungal, and virucidal capacities represents a breakthrough in public health and infection control. Inspired by natural defenses observed in the nano-scale architectures of insect wings and plant leaves, these biomimetic surfaces utilize mechano-biocidal mechanisms to target microbial and viral pathogens. The application of mechanical stress, accentuated by precise control of fluid dynamics around nanoengineered pillars, signifies a paradigm shift in mechanobiocidal strategies. Such approaches reduce reliance on pharmacological agents, mitigating the potential for emergent resistance and minimizing adverse biomedical implications.

Designing nanostructured surfaces with enhanced biocidal efficacy presents a formidable challenge at the intersection of multiple scientific disciplines. Insights gathered in this area emphasize the importance of meticulously tailoring surface features to optimize interactions with microorganisms. Researchers are progressively uncovering key design parameters and strategies crucial for enhancing the antimicrobial performance of these surfaces.

Structural parameters such as size, shape, and density of nanostructures are paramount. These features must be closely correlated with the dimensions and physical characteristics of the target pathogen to maximize physical disruption. Surface nanofeatures should be spaced appropriately to interact effectively with bacteria, fungi, or viruses, each requiring different levels of mechanical force for inactivation. Precisely patterned structures can rupture cell membranes in bacteria and fungi or damage protein coats of viruses, rendering them inactive.

The selection of materials for constructing nanostructured surfaces is another critical consideration. Metals such as silver, copper, and zinc are popular due to their recognized antimicrobial properties. However, polymers and carbon-based materials provide convenient fabrication methods and functional versatility. Combining the intrinsic biocidal properties of these materials with a carefully designed structure creates a composite effect that surpasses the capability of either strategy alone.

Understanding the dynamic interactions between surfaces and the environment has fueled interest in responsive and stimuli-reactive materials. These materials can alter their surface properties in response to environmental changes, such as shifts in pH, temperature, or the presence of microorganisms. This enables them to actively engage and inactivate pathogens in their vicinity.

Emerging interest centers around understanding the ecological and health impacts of nanostructured surfaces throughout their entire lifecycle. Prioritizing designs for longevity, safety, and environmental sustainability is crucial for broader acceptance and integration into public health frameworks. Future nanostructured surface designs will be guided by these critical insights, leveraging advanced fabrication techniques, computational modelling, and a deeper understanding of microbe-material interactions. Driven by innovation, these surfaces encompass a balance between form and function, balancing antimicrobial potency with practical and ecological considerations to deliver solutions that are as safe and sustainable as they are effective.

### 7.2. Challenges in Implementing Nanostructured Surfaces: Nanofabrication

Implementing nanostructured surfaces with biocidal properties, especially considering the intricacies of nanofabrication, poses considerable challenges. Creating features at the nanometer scale demands precision and control, pushing the limits of current technology and materials science[59]. Achieving the required resolution and pattern fidelity is a primary obstacle in nanofabrication[137]. Techniques like electron-beam lithography offer excellent precision but are time-consuming and not easily scalable for mass production. Similarly, nanoimprint lithography can replicate patterns with high throughput; however, the molds required are expensive and prone to wear, making the process unsuitable for all materials. Ensuring uniformity and consistency across large areas is another substantial challenge. Nanostructures must be evenly distributed and consistently shaped to maintain their biocidal efficacy, particularly for applications such as medical implants and devices, in which uniform surface properties are critical. Irregularities or defects in nanostructures can reduce antimicrobial performance or even encourage microbial colonization. The durability of nanostructures is also a concern, as regular wear and tear, including abrasion and exposure to cleaning agents, can degrade these features over time, compromising antimicrobial efficacy. This is particularly problematic in high-touch or high-traffic areas where frequent cleaning is necessary.

The economic viability of nanofabrication processes is a key consideration. The costs associated with high-precision techniques and materials used for biocidal nanostructures often exceed those of traditional antimicrobial treatments, potentially limiting the widespread adoption of these advanced surfaces. This challenge is particularly notable for public infrastructure and consumer goods, where cost-effectiveness is paramount.

Addressing these challenges requires continuous research and development to refine nanofabrication techniques and materials. Potential solutions include innovative approaches to reduce costs, enhance scalability, develop more robust materials and nanostructures, and improve our understanding of how these structures interact with microbes. Through collaborative efforts across disciplines, including engineering, microbiology, materials science, and manufacturing, the hurdles facing nanostructured surfaces can be overcome, leading to broader implementation and realization of their full potential for safeguarding against microbial threats.

Contemplating the inevitability of emerging microbial threats, the versatility and adaptability of these surfaces to counteract novel pathogens should be critically assessed. Future research must actively address these challenges by adapting and redefining the antimicrobial narrative to stay up to date with the evolving microbial environment.

The development of these advanced materials signifies a fundamental shift in tactics deployed to combat bacterial infections. This evolution represents a holistic, multidisciplinary approach that combines principles from chemistry, biology, and materials science to develop robust solutions. By mirroring the ingenious physical strategies employed by nature to ward off microbial threats, we not only forge a path for future breakthroughs but also address a pivotal challenge in the fight against infections, particularly in an era of escalating antibiotic resistance. As we venture forward, the prospects for bacterial management appear promising. Researchers are increasingly looking at nature's repository of designs and seeking inspiration to develop sophisticated materials. These materials are envisioned to provide durable, safe, and effective defence mechanisms against the minuscule yet formidable microbial adversaries that persist in the environment. This progressive approach promises to redefine our defence strategies against bacteria, offering hope against the backdrop of rising resistance and highlighting the power of integrating natural wisdom with scientific innovation.

## ASSOSIATED CONTENT

### Supporting information

The Supporting Information is available free of charge at

Supplemental information includes a Table 1S, which includes a summary of the specific effective antimicrobial rate on the surface of the biomimetic-designed nanostructures for bacterial cells, fungi and viruses

### Author Contributions

The manuscript was written through contributions of all authors. All authors have given approval to the final version of the manuscript. All authors contributed to the editing of the manuscript.

### Notes

The authors declare no competing financial interest.

## ACKNOWLEDGMENTS

V.A.B. acknowledge financial assistance from the Ministerio de Ciencia, Innovacion y Universidades of the Spanish Government through Research Project No. PID2020- 114347RB-C33, financed by No. MCIN/AEI 10.13039/ 501100011033. D.P.L. acknowledges funding from The University of Melbourne McKenzie Postdoctoral Fellowship Program. E.P.I. acknowledges support by the Australian Research Council by the ARC Research Hub for Australian Steel Manufacturing under the Industrial Transformation Research Hubs scheme (Grant ID No. IH130100017) and by the ARC Industrial Transformational Training (ITTC) Centre in Surface Engineering for Advanced Materials (SEAM) (Grant ID No. IC180100005). The authors acknowledge technical assistance by P.H. Le, S.W.I. Mah.